\documentclass[twocolumn,aps,floatfix,superscriptaddress]{revtex4}

\usepackage{graphicx,xcolor}
\usepackage{bm}

\usepackage{epsf}
\usepackage{amssymb}%
\usepackage{wrapfig}

\usepackage{amsmath,amsfonts}

\newcommand{\bea}{\begin{eqnarray}}
\newcommand{\eea}{\end{eqnarray}}
\newcommand{\beq}{\begin{equation}}
\newcommand{\eeq}{\end{equation}}
\newcommand{\be}{\begin{equation}}
\newcommand{\ee}{\end{equation}}

\newcommand{\p}{\partial}

\renewcommand{\>}{\rangle}
\newcommand{\<}{\langle}

\usepackage{color}

\usepackage{hyperref}

\begin{document}
\title{Gradient Catastrophe and Fermi Edge Resonances in Fermi Gases}

\author{E. Bettelheim}
\affiliation{Racah Institute of Physics, Hebrew University, Jerusalem, Israel}
\author{ Y. Kaplan}
\affiliation{Racah Institute of Physics, Hebrew University, Jerusalem, Israel}
\author{P. Wiegmann}
\affiliation{The James Franck  Institute, University of Chicago}
\begin{abstract}
Any smooth spatial  disturbance of a degenerate  Fermi gas inevitably becomes sharp. This phenomenon, called {\it the gradient catastrophe}, causes the breakdown of a Fermi sea to multi-connected components  characterized by multiple Fermi points. We argue that the  gradient catastrophe can be probed through a Fermi edge singularity measurement. In the regime of the gradient catastrophe the Fermi edge singularity problem becomes a non-equilibrium and non-stationary phenomenon. We show that the gradient catastrophe transforms  the single-peaked Fermi edge singularity of the tunneling (or absorption) spectrum to a sequence of multiple asymmetric singular resonances.  An extension of the bosonic representation of the electronic operator to non-equilibrium states captures the singular behavior of the resonances.
\end{abstract}
\maketitle
\noindent\paragraph*{1. Introduction} The FES  (Fermi edge singularity \cite{NozieresDedominicis,Mahan,Ohtaka:Tanabe}), observed as a power law peak in the absorption spectrum of X-rays in metals more than 70 years ago, is one of the most prominent and well understood quantum many-body phenomena caused solely by Fermi statistics.

FES also has been demonstrated in tunneling experiments \cite{Geim,Cobden,Hapke,IvanLarkin}:
a sudden switch-on of a contact potential due to a change in the capacity  of the contact  in tunneling causes a power law dependence of  the tunneling current on the bias voltage:  $I(V)\sim V^{-2a+ka^2}$ \cite{MatveevLarkin}.   Here $\delta=\pi a$ is the scattering phase of the ensuing potential and $k$ is the number of scattering channels.  In the case of an attractive potential $(a>0)$ the current  peaks at the Fermi edge.

The physics of the FES is explained by the phenomenon of the Orthogonality Catastrophe \cite{Anderson:Catastrophe}: the state of a Fermi gas $\<\Omega'|$ after a localized potential is suddenly switched on, is almost orthogonal to a state of the unperturbed  Fermi gas $|\Omega\>$.  Their overlap vanishes with the level spacing $\Delta$ as a power law $\<\Omega'|\Omega\>\sim(\Delta/E_F)^{ka^2}$.

\begin{figure}
\begin{center}
\vspace{-0.5cm}
\includegraphics[width=5cm]{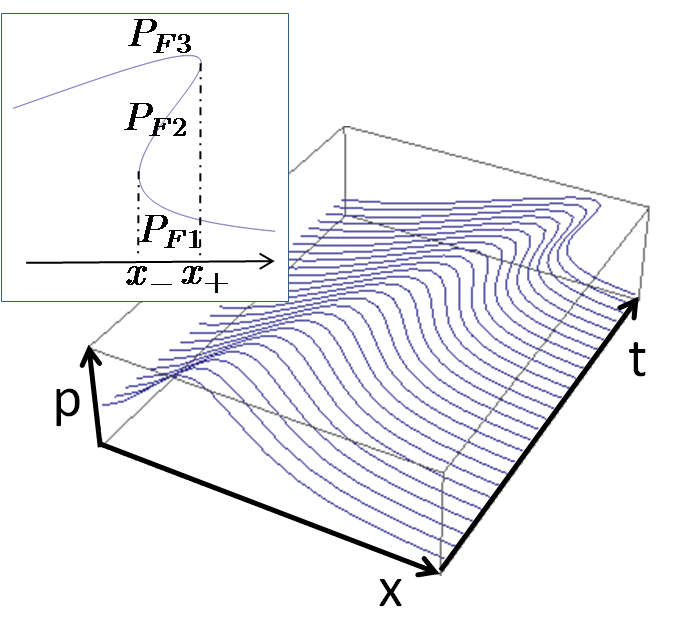}
\vspace{-0.5cm}
\caption{Time
progression of the Fermi edge.
The front  overhangs, giving rise to   three  Fermi (see inset) edges ($P_{F3}>P_{F2}>P_{F1}$)  between the trailing ($x_-$) and leading ($x_+$) edges. \label{1}}
\end{center}
\end{figure}

The FES  acquires new features in the non-stationary regime due to the  {\it gradient catastrophe}.  A gradient catastrophe  is a hydrodynamic instability observed in many classical (and, recently, in atomic) systems. In Ref. \cite{paperII} it has been shown that a quantum analog  of the gradient catastrophe also takes place in a degenerate Fermi gas. The ballistic propagation of  macroscopical packets and fronts in Fermi gases  inevitably enters the gradient catastrophe regime, where the initially smooth fronts develop large gradients and undergo a  shock wave phenomenon:  packets overturn as shown in Fig \ref{1} .
The observation of non-stationary  phenomena in Fermi gases may not be easy since electronic times are too short, but does not seem impossible.  From the theoretical viewpoint a non-stationary FES  reveals important (and new) aspects of  the Orthogonality Catastrophe. Both catastrophes are caused  solely by Fermi statistics and therefore their interaction is of interest.

The physics of non-stationary processes in Fermi gases is in its infancy. Ref. \cite{paperII}  discusses oscillatory corrections to the Orthogonality Catastrophe  in a non-stationary regime: the overlap of the state of a shaken-up Fermi gas  with a propagating packet (both before and after the shock). We also mention Refs. \cite{paper4} which cast non-stationary Fermi gases in the context of integrable non-linear waves.

\begin{figure}[b!!!]
\begin{center}
\includegraphics[width=5cm]{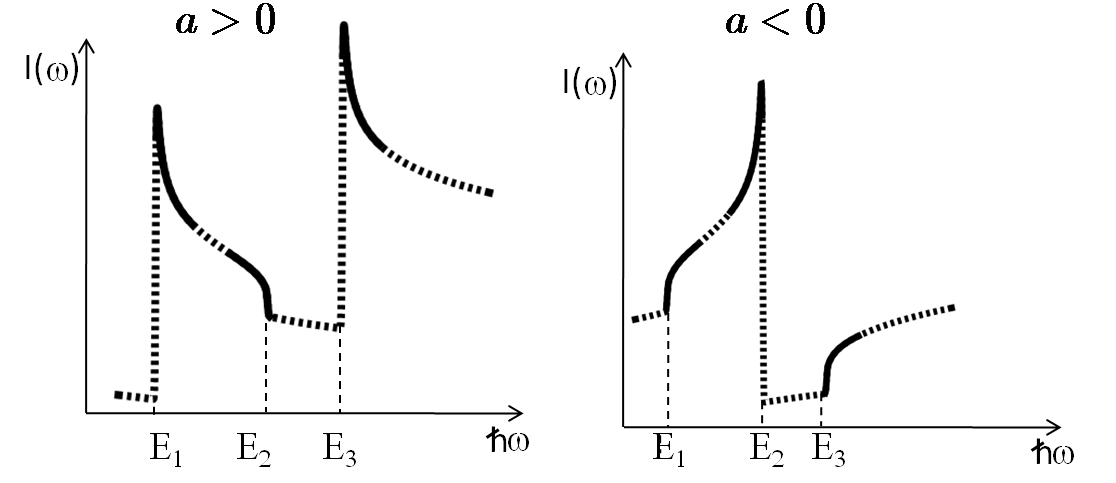}
\caption{A schematic plot of the tunneling current for  $a>0$ (left panel) and $a<0$ (right panel), solid lines show computed power law asymptotes.  Dashed lines interpolate between resonances. $a<0$ displays  a single peak after the shock, while $a>0$ displays two peaks.  \label{2}}
\end{center}
\end{figure}

In this paper we study the FES in a non-stationary regime before and after the quantum shocks had occurred.  We show that  each  shock introduces two  additional Fermi edges, each edge causes an  additional resonance peak schematically depicted in Fig.~\ref{2}. In general settings we predict that a smooth wave packet passing through a tunneling contact shows a sequence of singular pulses in the tunneling current. Here we discuss only one shock, though the extension to multiple shocks is straightforward.

Non-stationary FES in its generality consists of several different regimes some of them are  difficult to study. In this paper we consider a regime away from turning points. There the Fermi surface changes much slower than the Fermi scale. In this case the problem becomes quasi-stationary and can be solved in two steps. First we neglect the time dependence of the Fermi points leading to  FES problem with a non-equilibrium population of electronic states. Once this has been done, we may  treat the Fermi-edges as slow time dependent variables. We argue that they obey Riemann equation.

The stationary non-equilibrium FES problem has been  extensively studied in Refs. \cite{Combescot} and more recently in Refs.  \cite{Levitov:Abanin, Mirlin:Gutman:Gefen, BKW}, so we could have borrowed the results. We suggest instead a simple compact approach to the non-equilibrium FES  which captures the singular parts of the resonances. This approach extends the representation of the electronic operator by means of Bose fields to non-equilibrium states. In a related paper \cite{BKW} we  formulate the non-equilibrium FES problem as a matrix Riemann-Hilbert problem with an integrable kernel.

\paragraph*{2. Tunneling in a non-stationary regime.} We consider the  following situation:\\
\noindent (i) A Fermi gas is in contact  with  a localized resonant level (a quantum dot). It is initially uncharged and provides no scattering to electrons.
When an electron tunnels to the dot, it  suddenly charges the dot, switching-on a  small potential $H \to H'=H+U$ localized at the dot \cite{MatveevLarkin}. We assume the potential to be weak, such that the scattering phase $|\delta|<\pi/2$ and therefore $|a|<1/2$. Assume  no further  interaction, no dissipation, ignore spin and channels.\\
\noindent (ii) A semiclassical electronic front or a packet  - a state with a spatially inhomogeneous density matrix $\varrho$  has been initially created in a Fermi gas  as it is shown in  Fig. \ref{1} \cite{paperII}.

This setting can be realized in different ways. For example, a smooth potential well (namely, of spatial extent much larger than the Fermi length) centered away from the dot is applied to the Fermi gas. A large number of electrons is trapped in the well. Then the well is suddenly removed. The electronic packet propagates towards the dot,  eventually hits it, facilitating tunneling. We may consider two conductors with different chemical potentials suddenly brought into contact. The conductor with smaller chemical potential is in contact with a dot. A front of electronic density propagates toward the smaller chemical potential  of the dot.

Alternatively one may  apply a time dependent voltage through a point like contact separated from the dot \cite{Levitov:Klich}.

A state created in such processes is a Fermi coherent state
$$|\Omega\>=\exp\left(\frac{i}{\hbar}\int P_0(x)\varphi(x)dx\right)|0\>$$
obtained by a unitary rotation of the ground state $|0\>$  of a Fermi gas.  A function $P_0(x)\ll  P_F$ is an initial momenta of the packet.
The field $\varphi$ is a chiral canonical Bose field related to the chiral part of electronic density
\begin{equation}\label{B}
\varphi(x)=\hbar\sum_{k\neq0} \frac 1k e^{\frac{i}{\hbar}kx}\rho_k, \quad \rho_k=\sum_{p} c^\dag_{p}c_{p+k},
\end{equation}
where $p$ and $p+k$ are electronic momenta close to the Fermi momentum $P_F$.
We assume that $P_0(x)$ is smooth  on the Fermi length scale $\hbar |\nabla P_0|/P_0\ll P_F$.  This condition justifies the semiclassical analysis described below.

The tunneling current is given by the golden rule \cite{NozieresDedominicis,MatveevLarkin}. In units of a tunneling amplitude  $I(\omega)|_{\hbar\omega=eV+E_F}$  reads
\begin{align}
 &I(\omega,t)\propto \mbox{Re}  \int_0^\infty e^{i\omega \tau}   G(t+\frac{\tau}{2},t-\frac{\tau}{2}) d\tau,\\
 \label{red}
&G(t_1,t_2) =  \<\Omega|e^{iHt_2}c e^{iH'(t_{1}-t_2)}
 c^\dag\; e^{-iHt_1}|\Omega\>
\end{align}
Here $c=\sum_k e^{-ikx_0}c_k$ and $x_0$ is the position of the dot.

Since all the physics is concentrated at the Fermi edge a knowledge of the dispersion at the edge is sufficient
\begin{align}
 \label{e}\epsilon_p=E_F+v_F(p-P_F)+\frac{(p-P_F)^2}{2m}+\dots.
\end{align}
In the literature the parabolic part of the dispersion is routinely ignored. In this approximation our effect disappears \cite{Glazman}.

We evaluate $G(t_1,t_2)$ in the regime where the typical time of tunneling $\tau=t_2-t_1>0$ is much smaller than the time it takes for the packet to change.  This approximation does not allow us to compute the broadening of the resonance at the frequency range $ \gamma\sim v_F(\nabla P_0/\hbar)^{1/2}$, but it  captures the power law shoulders of resonances and their dependence on time $t$.

Under this assumption during the short time of tunneling the energy dependence of the Fermi velocity and  scattering phase $\delta$ caused by the potential $U$ can be dropped in some  interval $|\epsilon -E_F|\ll \Lambda$ at the Fermi edge, where the cut-off $\Lambda$ (typically of the Fermi scale) is assumed to be  larger than $v_F P_0$ and $\hbar/\tau$. This amounts a shift of energy levels after scattering downwards by a constant amount  $a$ (in units of level spacing):  $\epsilon_p\to \epsilon_p-a$.

In Ref. \cite{SchotteSchotte} it has been shown that  the vertex operator $e^{a\varphi}$ implements a  shift of momenta  such that a perturbed Hamiltonian and perturbed states are
$H' =  e^{-a \varphi(x_0) } H e^{ a\varphi(x_0)}$ and $|\Omega'\>=e^{a\varphi(x_0)}|\Omega\>$. Then Green's function reads
\begin{align} \label{G}
G(t_1,t_2) =  \<\Omega|c(t_2)
e^{ - a\varphi(x_0,t_2) }e^{  a\varphi(x_0,t_1)} c^\dag(t_1)|\Omega\>.
\end{align}
This formula is standard.  The only difference is that the density matrix  does  not commute with the Hamiltonian and therefore the process  is not-stationary - the Green function and a current depend on $t=1/2(t_1+t_2)$.

\paragraph*{3. Gradient Catastrophe: Riemann equation for Fermi Gases.}
We demonstrate the gradient catastrophe  on the evolution  of the Wigner function - a simpler  object than (\ref{red}). The Wigner function describes occupation in phase space:
\begin{align}\label{Wignerdefinition}
n_F(x,p,t) = \int \<\Omega| c^\dagger(x +\frac y2,t) c( x - \frac y2,t)
|\Omega\> e^{ -\frac{i}{ \hbar}py } dy
\end{align}
We assume that the front is plane or radial, such that the dynamics is essentially one dimensional and chiral.

Semiclassically, the Wigner function is equal to 1 in a bounded domain $p<P_F(x,t)$ of the phase space $(p,x)$ - the Fermi sea - and vanishes outside the Fermi sea $n_F(x,p,t)\approx\Theta(P_F(x,t)-p)$.  This form is valid as long as the gradients of the spatial dependence of the Fermi momentum   $P_F(x,t)$ are small \cite{BW}.  The shape of the initial  Fermi surface  is given by the density matrix  $P_F(x,0)=P_F+P_0(x)$. The support of the Wigner function is the area below the Fermi surface in Fig.~\ref{1}.

How does the Fermi surface change in time? It does not, if one neglects dispersion of the Fermi gas, i.e., treats the velocity $v_p=d \epsilon_p/dp$ as a constant: the front  translates with the Fermi velocity   without changing its shape. It does change, in  a dramatic fashion,  if the dispersion  in (\ref{e}) (no matter how small) is taken into account.

The Wigner function (for a dispersion $\epsilon_p=p^2/2m$) obeys the equation
\begin{align}
\left(\partial_t  +v_p  \nabla  \right)n_F(x,p,t)=0, \quad v_p=p/ m
 \end{align}
The solution of this equation
\begin{equation}
n_F(x,p,t)= n_F(x-v_pt,p,0)\approx \Theta\left(P_0(x-\frac pm t)-p\right)
\end{equation}
shows that  a moving Fermi momentum $P_F(x,t)$ obeys a {\it hodograph} equation
\begin{align}\label{h}
P_F(x,t) = P_0\left( x - P_F(x,t)/m\cdot t\right)
\end{align}
This is Riemann's solution of Euler's equations for hydrodynamics of a compressible one-dimensional fluid, also called Riemann (or Riemann-Burgers, or Riemann-Hopf) equation
\begin{align}\label{Riemann}
\p_tP_F+\nabla E_F=0,\quad E_F(x,t)={P_F^2(x,t)}/{2 m}
\end{align}
The Riemann equation leads to shock waves: the velocity of a point with momentum $P_F(x)$  is $P_F(x)/m$:  higher parts of the front move faster. The front  gets steeper, and eventually attains an infinite gradient -- a shock  at some finite time. After this moment the Riemann equation has at least three real solutions, $P_{F3}(x,t)\!\!>\!\!P_{F2}(x,t)\!>\! P_{F1}(x,t)$, confined between two turning points $x_-(t),x_+(t)$, the trailing and  leading edges respectively (Fig.\ref{1}).

This phenomena is the gradient catastrophe. Any smooth disturbance of the Fermi surface eventually arrives to a point where the Fermi surface acquires infinite gradients, and then becomes multi-valued between moving turning points $x_\pm$. We focus on that region, namely the region where all electronic states with energy below $E_{F1}$ and between $E_{F2}$ and $E_{F3}$ are occupied. The Fermi distribution acquires at least three, or more edges.

\paragraph*{5. Slowly evolving Fermi edges}
Away from turning points   we can employ the  Whitham averaging method known in the theory of non-linear waves \cite{Whitham}. The method has been applied  to electronic systems in \cite{paperII}. It  is based on a separation of scales between the slowly varying Fermi points and fast oscillations of the electronic states. In short, the  Whitham method suggests treating  the slowly changing Fermi edges as constants while computing Green's function (\ref{red}), and  then to include the motion of the Fermi edges  in the final result.  Motion of the Fermi edges is determined by the Riemann equation (\ref{Riemann}). This approach is valid away from turning points \cite{comment3}). It can be justified mathematically using an integrable non-linear equation for Green's function obtained in \cite{paper4}. We omit the mathematical justification of this procedure since the approximation of slowly moving Fermi-edges  is physically obvious.

\begin{figure}[h!!!]
\begin{center}
\includegraphics[width=4cm]{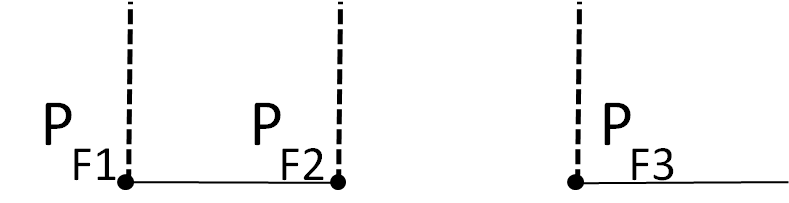}
\caption{Unoccupied electronic states are between edges  $E_{F1}$ and $E_{F2}$ and above $E_{F3}$. \label{Contour}}
\end{center}
\end{figure}

In the shock region $x_-<x_0<x_+$ we must  evaluate the current  (\ref{red}) over a state with three Fermi edges, where electrons occupy states below $E_{F1}$ and between $E_{F2}$ and $E_{F3}$  (see Fig.\ref{Contour}), and then treat the time dependent edges as solutions of the hodograph equation (\ref{h}).  This is a non-equilibrium (and stationary FES problem) addressed in  Refs. \cite{Combescot,Levitov:Abanin,Mirlin:Gutman:Gefen, BKW}.
\paragraph*{6.  Fermi edge resonances} Now we are ready to present our result: the current at a frequency close to the edges  $\hbar\gamma\ll \varepsilon_i\left(\hbar\omega-E_{Fi} \right) \ll v_F P_0 $, in  units  of  cut-offs reads:
\begin{align}\label{result1}
I(\omega,t)\propto \left| \hbar\omega-E_{Fi}\right |^{3a^2-2\varepsilon_ia}\prod_{j\neq i}E_{ij}^{-2\varepsilon_j a} \prod_{n> m} E_{nm}^{\,2\varepsilon_n\varepsilon_m a^2},
\end{align}
where  $\varepsilon_i=(-1)^{i+1},\,i=1,2,3$ and $E_{ij}=E_{Fi}-E_{Fj}$. The Fermi energies all depend on space- time according to the hodograph equation (\ref{h}). We suppress this dependence in the notations.

The last factor in this equation gives an overlap between states before and after the shake-up - an  Orthogonality Catastrophe formula for multiple edges
\begin{align}
|\<\Omega|\Omega'\>|=\Delta^{3a^2}\prod_{n>m}E_{nm}^{\epsilon_n\epsilon_m a^2},
\end{align}
where $\Delta$ is a level spacing. It is a power of  Cauchy determinant $\mbox{det}\frac{\Delta}{E_{2i+1}-E_{2j}}=\Delta^3\prod_{n>m}E_{nm}^{\epsilon_n\epsilon_m }$ constructed out of lower and upper edges of occupied bands. These formulas extend the result in \cite{NozieresDedominicis,Mahan,Ohtaka:Tanabe} to multiple edges (see also \cite{Combescot,Levitov:Abanin,Mirlin:Gutman:Gefen, BKW}).

If the potential is attractive $a>0$ (a common case) the current features a peak at $E_{F1}$  (almost zero bias) and an additional  resonance at  $E_{F3}$ with a power law to the right of the edges. Current is suppressed at the edge $E_{F2}$. If potential is repulsive $a<0$ a peak may appear at the edge $E_{F2}$ with a power law to the left to the edge  Fig.\ref{2}.

Apart from additional resonances, the unique features of the shock region is the value of the exponent and the time-dependent  amplitudes.
Outside of the shock region where there is only one Fermi edge or at a larger energy $ \left |E_{Fi}-E_{Fj}\right |\ll \left |\omega- E_F\right |\ll\Lambda$ where the fine structure of Fermi edges becomes negligible, the current is given by the standard formula \cite{NozieresDedominicis,Mahan} $I(\omega)\propto
\left( \omega-E_{F}\right)^{a^2-2a}$ where only the edge $E_F$ depends on time.
\paragraph*{7. Bosonic representation for non-equilibrium states} Techniques developed in Refs. \cite{Combescot,Levitov:Abanin,Mirlin:Gutman:Gefen, BKW} allow computing the details  of the FES resonances. However the most interesting singular power law  asymptote next to the edges can be captured by a simpler approach which extends the familiar representation of electronic operators through Bose fields \cite{SchotteSchotte}.  We briefly discuss it below.

First we separate fast oscillatory modes at each edge  $$c(t,x_0)=\sum_i e^{\frac{i}{\hbar}P_{Fi}x(t)} \psi_i(t)$$ where $\psi_i(t)$ are slowly changing modes and $x(t)=x_0-v_Ft$. Then we represent each slow  modes  through components of the Bose field as
\begin{align}\label{FermionOperator}
\psi_i\propto (\varepsilon_i \prod_{j\neq i}E_{ij}^{\varepsilon_j})^{1/2}e^{-\varepsilon _i \varphi_i}.
\end{align}
Under this representation gradients of components of the Bose field
\begin{align}\label{B}\partial_x \varphi_i = i \psi^\dag_i(t)\psi_i(t), \quad  \varphi=\sum_i\varphi_i.
\end{align}
are density of slow electronic modes. The Bose field is a sum of its components. The origin of the important factor in front of $e^{-\varepsilon _i \varphi_i}$ will be clear in a moment.

The component of the Bose field represent particle-holes excitations close to each edge. At $\tau E_{ij}\gg\hbar  $  they can be treated as independent canonical Bose fields. Their variances $C_i(t_1,t_2)=-\frac 12\<\Omega|\left(\varphi_i(t_2)-\varphi_i(t_1)\right)^2|\Omega\>$   are not difficult  to compute. As follows form (\ref{B}): $C_i$ are electronic density  correlations at each edge:, $C_1$ and $C_3$  are sums of  $(\cos\epsilon\tau-1)/\epsilon$ over all possible energies of particle-hole excitations provided that a particle is taken out from the first band $\epsilon<E_{F1}$ and the second band $E_{F2}<\epsilon<E_{F3}$ respectively. Similarly $C_2$ is  the sum over momentum of a hole particle  excitations provided that a hole is taken out from  the  "gap" between $E_{F1}$ and $E_{F2}$.  For example $C_2=\left(\sum_{\epsilon>0}^{E_{21}}-\sum_{E_{32}}^\Lambda\right)(\cos\epsilon\tau-1)/\epsilon$. Computing these integrals at $\tau \gg\hbar/|E_{ij}|$ one obtains
\begin{align}\label{CC}
C_i(\tau)=-\log\tau+\varepsilon_i \sum_{j\neq i}{\varepsilon_{j}}\log |E_{ij}|
\end{align}
The $\tau$-independent term in (\ref{CC}) (a zero mode of the Bose field) explains the  pre-factor in (\ref{FermionOperator}). Together  (\ref{FermionOperator}) and (\ref{CC}) produce a canonical electronic  correlation function
\begin{align}\label{psi}\<\Omega|\psi^\dag_i(t_1)\psi_i(t_2)|\Omega\>\propto
\frac{\varepsilon_i}{x(\tau)}\left( \prod_{j\neq i}E_{ij}^{-\varepsilon_i \varepsilon_j}\right)\,e^{C_i}.
 \end{align}
In the Bose representation, Green's function (\ref{G}) is a sum of edge components
\begin{align}\label{G0}
 &G(t_1,t_2)=\sum_i \left(\varepsilon_i \prod_{j\neq i}E_{ij}^{-\varepsilon_i \varepsilon_j}\right)e^{\frac i\hbar E_{Fi}\tau}G_i(t_1,t_2),\\
 \nonumber
 &G_i=\<e^{(\varepsilon_i-a)(\varphi_i(t_2)-\varphi_i(t_1))}\>
\prod_{j\neq i}\<e^{-a(\varphi_j(t_2)-\varphi_j(t_1))}\>=\\
&=e^{(\varepsilon_i-a)^2 C_i} e^{a^2\sum_{j\neq i}C_j}=e^{(1-2\varepsilon_ia) C_i}e^C .
 \end{align}
We obtain Green's function in the form of  two  factors: 'closed loops' $e^C$  where each edge contributes equally and 'open lines'  $L_i$ corresponding to each edge \cite{NozieresDedominicis}.
 \begin{align}\label{CL}
&G=e^C\cdot L,\quad C=a^2\sum_i  C_i,\quad L=\sum_i L_i,\\
&L_i=\varepsilon_i e^{\frac i\hbar E_{Fi}\tau}\prod_{j\neq i}E_{ij}^{{{-\varepsilon_j\varepsilon_i}}}e^{(1-2\varepsilon_ia)C_i}, \\
&C=-3a^2\log\tau+a^2 \sum_{j\neq i}\varepsilon_i\varepsilon_{j}\log |E_{ij}|
 \end{align}
This prompts  the formula for the current (\ref{result1}).

 {\it Acknowledgment} The authors acknowledge discussions with A. Abanov on all aspects of this work. P. W. was supported by NSF DMR-0906427, MRSEC under DMR-0820054. E. B. was supported by grant 206/07 from the ISF.


\begin{thebibliography}{10}
\bibitem{comment3} At turning points where gradients are of the order of a Fermi scale  enhanced quantum corrections blur  Fermi edges \cite{BW}.
\end{thebibliography}
\end{document}